# Analyzing Signal Attenuation in PFG Anomalous Diffusion via a Modified Gaussian Phase Distribution Approximation Based on Fractal Derivative Model

Guoxing Lin*

Carlson School of Chemistry and Biochemistry, Clark University, Worcester, MA 01610

ABSTRACT

Pulsed field gradient (PFG) technique is a noninvasive tool, and has been increasingly employed to study anomalous diffusions in Nuclear Magnetic Resonance (NMR) and Magnetic Resonance Imaging (MRI). However, the analysis of PFG anomalous diffusion is much more complicated than normal diffusion. In this paper, a fractal derivative model based modified Gaussian phase distribution method is proposed to describe PFG anomalous diffusion. By using the phase distribution obtained from the effective phase shift diffusion method based on fractal derivatives, and employing some of the traditional Gaussian phase distribution approximation techniques, a general signal attenuation expression for free fractional diffusion is derived. This expression describes a stretched exponential function based attenuation, which is distinct from both the exponential attenuation for normal diffusion obtained from conventional Gaussian phase distribution approximation, and the Mittag-Leffler function based attenuation for anomalous diffusion obtained from fractional derivative. The obtained signal attenuation expression can analyze the finite gradient pulse width (FGPW) effect. Additionally, it can generally be applied to all three types of PFG fractional diffusions classified based on time derivative order $\alpha$ and space derivative order $\beta$. These three types of fractional diffusions include time-fractional diffusion with $\{0 < \alpha \leq 2, \beta = 2\}$, space-fractional diffusion with $\{\alpha = 1, 0 < \beta \leq 2\}$, and general fractional diffusion with $\{0 < \alpha, \beta \leq 2\}$. The results in this paper are consistent with reported results based on effective phase shift diffusion equation method and instantaneous signal attenuation method. This method provides a new, convenient approximation formalism for analyzing PFG anomalous diffusion experiments. The expression that can simultaneously interpret general fractional diffusion and FGPW effect could be especially important in PFG MRI, where the narrow gradient pulse limit cannot be satisfied.

*Address:Department of Chemistry, Clark University,950 Main St, Worcester, MA 01610, Fax:+1 (508)793-7117, Email:glin@clarku.edu

## 1. Introduction

Anomalous dynamical behavior [1, 2, 3,4, 5, 6,7] exists in many systems. Many experimental and theoretical efforts have been put in investigating non-Gaussian diffusion behavior in pulsed field gradient (PFG) Nuclear Magnetic Resonance (NMR) and Magnetic Resonance Imaging (MRI) [8, 9, 10, 11]. These efforts include the propagator representation [12], Gaussian phase distribution (GPD) approximation [13, 14], short gradient pulse (SGP) approximation [15], the stretched exponential models [16, 17,18, 19], the walk and spectral dimension parameters method [20], the modified Bloch equations [21,22,23], the log-normal distribution function [24], the effective phase shift diffusion equation (EPSDE) [25], and the instantaneous signal attenuation method [26]. However, PFG anomalous diffusion research is still in its early stages due to the difficulties in the theory and the complexity of the application systems [27,28]. Therefore, it is still important to develop new theoretical treatments for anomalous diffusion.

It is particularly difficult to analyze PFG signal attenuation inside each gradient pulse, namely the finite gradient pulse width (FGPW) effect, in the case of a general fractional diffusion with time derivative parameter $\alpha$ and space derivative parameter $\beta$ belonging to $\{0 < \alpha, \beta \leq 2\}$. Understanding FGPW effect is important, since using a longer gradient pulse can measure a slower diffusion under the same maximum gradient strength, and especially in PFG MRI, the narrow gradient pulse limit approximation often cannot be applied. Until recently, two different theoretical signal attenuation expressions including the FGPW effect for general fractional diffusion with $\{0 < \alpha, \beta \leq 2\}$ were obtained [25,26], which are $\exp\left[-D_{f_1}\int_0^\tau K^\beta(t)dt^\alpha\right]$ and $E_\alpha\left[-D_{f_2}\int_0^\tau K^\beta(t)dt^\alpha\right]$, based on the fractal derivative model [29,30] and the fractional derivative model [31,32], respectively, where $E_\alpha$ is a Mittag-Leffler function, $K(t)$ is the wavenumber defined by $K(t) = \int_0^t \gamma g(t')dt'$, and $D_{f_1}$ and $D_{f_2}$ are the fractional diffusion coefficients with units of $m^\beta/s^\alpha$. In references [25], it has shown that these two expressions demonstrate complementary nature to each other. Since the available expressions including the FGPW effect for general fractional diffusion are very few and may be inconvenient such as the integration inside these two expressions mentioned above, it may provide new insight to develop some new theoretical methods. Such new approaches may be developed by modifying the traditional GPD approximation method, and using a phase distribution that is obtained from the fractional diffusion equation. Two different types of phase distributions for fractional diffusion have recently been obtained by the effective phase shift diffusion equation method [25] based on fractional derivatives and fractal derivatives. One type of phase distribution obtained based on the fractional derivative model has been used to derive a Mittag-Leffler function based PFG signal attenuation expression that includes the FGPW effect for general fractional diffusion [33]. Another type of phase distribution based on the fractal derivative model could also be important, since other methods [25,26] have demonstrated that the fractal derivative model gives a stretched exponential function based signal attenuation. This kind of attenuation formalism based on stretched exponential function has been used to fit PFG diffusion in polymer and biological systems [18,19,25,26,34,35]. Additionally, an empirical stretched exponential function based attenuation expression has been used to study water diffusion in brain tissue using MRI 16,17]. Therefore, in this paper, the phase distribution based on the fractal derivative model is employed to derive a new PFG signal attenuation expression. This derivation employs techniques and results from the conventional GPD approximation such as Kärger et al.'s time-correlation function calculation results (see Appendix A) [13]. The signal attenuation expression obtained in this paper includes the FGPW effect for all three types of fractional diffusion: general fractional diffusion $\{0 < \alpha, \beta \leq 2\}$, time fractional diffusion $\{0 < \alpha \leq 2, \beta = 2\}$ and space fractional diffusion $\{\alpha = 1, 0 < \beta \leq 2\}$ [25,31,32]. The results obtained by this method agree with other results in the literature within a certain range and provide a new way to interpret PFG fractional diffusion.

The rest of this paper is organized as follows. In section 2, from the fractional diffusion equation based on the fractal derivative model, the expression of the mean $\beta$-th power of the phase displacement is derived, and then used to obtain the modified phase distribution which determines the signal attenuation of PFG anomalous diffusion. In section 3, the results and discussion section, the results are compared to literature results, and the differences between the fractal derivative and fractional derivative models are discussed. Additionally, the general expressions for determining the time derivative and space derivative parameters based on the obtained signal attenuation expression are discussed.

2. **Theory**

The Hausdorff derivative has been employed by Chen to derive an anomalous transport diffusion equation [29]. The fractal derivatives are defined by [29,30]

$$\frac{\partial P^\beta}{\partial t^\alpha} = \lim_{t_1 \to t} \frac{P^\beta(t_1) - P^\beta(t)}{t_1^\alpha - t^\alpha}, 0 < \alpha, 0 < \beta, \tag{1}$$

and the fractional diffusion can be described as [29,30]

$$\frac{\partial P(z,t)}{\partial t^\alpha} = D_{f_1} \frac{\partial}{\partial z^{\beta/2}} \left( \frac{\partial P(z,t)}{\partial z^{\beta/2}} \right), \tag{2}$$

where $z$ is the position, $t$ is the time, and $P(z,t)$ is the probability distribution function (PDF) in real space. This equation has been employed to study anomalous diffusion processes such as water diffusion in unsaturated media [36]. It has also been used to obtain the PFG signal attenuation expression $\exp\left[-D_{f_1}\int_0^\tau K^\beta(t)dt^\alpha\right]$ [25,26], which is consistent with other literature results [13, 21, 35] such as Magin et al.'s result of space fractional diffusion [21]. This expression and the consistently theoretical results in literature [13, 21, 35] have been applied to analyze the PFG signal attenuation data in polymer and biological systems [21, 25, 26, 34, 35]. Here, it will be shown how a new approximation signal attenuation expression can be obtained based on the equation.

By introducing $z' = z^{\beta/2}$ [29] and $t' = t^\alpha$, Eq. (2) can be transformed into

$$\frac{\partial P'(z',t')}{\partial t'} = D_{f_1} \frac{\partial}{\partial z'}\left(\frac{\partial P'(z',t')}{\partial z'}\right), \tag{3}$$

where $P'(z',t')$ is the PDF in $z'$ variable space. Eq. (3) has the same form as the normal diffusion equation and its solution is [29]

$$P'(z',t') = \frac{1}{\sqrt{4\pi D_{f_1} t'}} \exp\left(-\frac{z'^2}{4D_{f_1} t'}\right). \tag{4}$$

$P'(z',t')$ satisfies the normalization condition

$$\int_{-\infty}^{\infty} P'(z',t')dz' = 1. \tag{5}$$

By substituting $z' = z^{\beta/2}$ and $t' = t^\alpha$ into Eq. (5), we have

$$\int_{-\infty}^{\infty} P(z',t')dz' = 2\int_{0}^{\infty} \frac{1}{\sqrt{4\pi D_{f_1} t^\alpha}} \exp(-\frac{z^\beta}{4D_{f_1} t^\alpha})dz^{\frac{\beta}{2}} = 2\int_{0}^{\infty} \frac{\beta z^{\frac{\beta}{2}-1}}{2\sqrt{4\pi D_{f_1} t^\alpha}} \exp(-\frac{z^\beta}{4D_{f_1} t^\alpha})dz. \qquad (6)$$

From Eq. (6), the normalized probability distribution function $P(z,t)$ can be obtained as

$$P(z,t) = P'(z',t)\left|\frac{dz'}{dz}\right| = \frac{\beta |z|^{\frac{\beta}{2}-1}}{2\sqrt{4\pi D_{f_1} t^\alpha}} \exp(-\frac{|z|^\beta}{4D_{f_1} t^\alpha}), \qquad (7)$$

which is an even function. The mean $\eta$-th power of the displacement $\langle |z^\eta(t)| \rangle$ can be calculated by

$$\langle |z^\eta(t)| \rangle = \int_{-\infty}^{\infty} |z^\eta| P(z,t) dz = \frac{1}{\sqrt{\pi}} (4D_{f_1} t^\alpha)^{\frac{\eta}{\beta}} \Gamma\left(\frac{\eta}{\beta}+\frac{1}{2}\right). \qquad (8)$$

Another way to obtain $\langle |z^\eta(t)| \rangle$ is to calculate directly in $z'$ variable space. As $z = z'^{2/\beta}$, $\langle |z^\eta(t)| \rangle$ can be calculated as

$$\langle |z^\eta(t)| \rangle = \int_{-\infty}^{\infty} |z'^{2\eta/\beta}| P(z',t) dz' = \frac{1}{\sqrt{\pi}} (4D_{f_1} t^\alpha)^{\frac{\eta}{\beta}} \Gamma\left(\frac{\eta}{\beta}+\frac{1}{2}\right), \qquad (9)$$

which is exactly the same as that obtained in Eq. (8). From Eq. (8), we have

$$\langle |z^\beta(t)| \rangle = 2D_{f_1} t^\alpha, \qquad (10)$$

which agrees with W. Chen's result in reference [29] and

$$\langle z^2(t) \rangle = \frac{1}{\sqrt{\pi}} (4D_{f_1} t^\alpha)^{\frac{2}{\beta}} \Gamma\left(\frac{2}{\beta}+\frac{1}{2}\right). \qquad (11)$$

which has the same time dependence behavior of $\langle z^2(t) \rangle \propto t^{\frac{2\alpha}{\beta}}$ as that obtained in fractional derivative model [31,32].

In a PFG diffusion experiment, the accumulating phase shift $\phi(t)$ of the spin carriers affected by the gradient pulses in a rotating frame can be described as [13, 35, 37]

$$\phi(t) = \int_{0}^{t} \gamma g(t') z(t') dt', \qquad (12)$$

where $\gamma$ is the gyromagnetic ratio, $g(t')$ is the gradient strength at time $t'$, and position $z(t')$ is a time-dependent variable related to the spin self-diffusion process. In the effective phase shift diffusion equation method, the accumulating phase shift can be described using a random walk model as [25, 38]

$$\phi(t_{tot}) = \begin{cases} \sum_{\Delta t}[K(t_{tot})-K(t)]\Delta z(t) + K(t_{tot})z_0, & K(t_{tot}) \neq 0 \\ -\sum_{\Delta t} K(t)\Delta z(t), & K(t_{tot}) = 0 \end{cases}, \qquad (13)$$

where $K(t)$ is the wavenumber defined by $K(t) = \int_0^t \gamma g(t')dt'$, $z_0$ is the starting position, $t_{tot} = \sum \Delta t$ is the time at the end of the refocusing gradient pulse, and $\Delta z(t)$ is the jump length during $\Delta t$ at time $t$. Usually, $K(t_{tot}) = 0$, but $K(t_{tot}) \neq 0$ can exist in some intermolecular multiple quantum coherence (iMQC) sequences [38]. In Eq. (13), the summation $\sum_{\Delta t} K(t)\Delta z(t)$ can be treated as a diffusion process in phase space, since it is similar to the diffusion process of spin carriers in the real space that can be described as $\sum_{\Delta t} \Delta z(t)$. There is a scaling factor $K(t)$ between the lengths of corresponding jumps of these two diffusion processes in phase space and real space, which leads to an effective diffusion coefficient $D_{\phi eff,f_1}(t) = K^\beta(t) D_{f_1}$ with units of rad$^\beta$/s$^\alpha$ since $D_{\phi eff,f_1}(t)/D_{f_1} \propto [K(t)\Delta z(t)]^\beta / [\Delta z(t)]^\beta$ [25]. As the two diffusions belong to the same type of diffusion, they should obey the same type of diffusion equations. From Eq. (2), by replacing the diffusion coefficient $D_{f_1}$ with $D_{\phi eff,f_1}(t)$, and coordinate parameter $z$ with $\phi$, the fractional diffusion equation with a time-dependent diffusion constant $D_{\phi eff,f_1}(t)$ can be obtained as [25]

$$\frac{\partial P(\phi,t)}{\partial t^\alpha} = D_{\phi eff,f_1}(t) \frac{\partial}{\partial \phi^{\beta/2}}(\frac{\partial P(\phi,t)}{\partial \phi^{\beta/2}}), \qquad (14)$$

where $P(\phi,t)$ is the PDF in phase space. Under the SGP approximation, the obtained phase shift is similar to Eq. (7) [25]

$$P(\phi,t) = \frac{\beta |\phi|^{\frac{\beta}{2}-1}}{2\sqrt{4\pi D_{\phi eff,f_1}(\delta)t^\alpha}} \exp(-\frac{|\phi|^\beta}{4 D_{\phi eff,f_1}(\delta)t^\alpha}), \qquad (15)$$

The mean $\eta$-th power of the displacement $\langle |\phi^\eta(t)| \rangle$ can be calculated as

$$\langle |\phi^\eta(t)| \rangle = \frac{1}{\sqrt{\pi}}(\gamma g \delta)^\eta (4 D_{f_1} t^\alpha)^{\frac{\eta}{\beta}} \Gamma\left(\frac{\eta}{\beta} + \frac{1}{2}\right). \qquad (16)$$

From Eq. (16), we have

$$\frac{\langle |\phi^{\beta}(t)| \rangle}{2} = \frac{1}{4} \left[ \frac{\langle \phi^{2}(t) \rangle}{\frac{1}{\sqrt{\pi}} \Gamma\left(\frac{2}{\beta} + \frac{1}{2}\right)} \right]^{\frac{\beta}{2}}. \tag{17}$$

And Eq. (15) can be rewritten as

$$P(\phi, t) = \frac{\beta |\phi|^{\frac{\beta}{2}-1}}{2\sqrt{2\pi \langle |\phi^{\beta}(t)| \rangle}} \exp\left(-\frac{|\phi|^{\beta}}{2 \langle |\phi^{\beta}(t)| \rangle}\right). \tag{18}$$

The signal attenuation has been obtained by the effective phase shift diffusion equation method and is [25]

$$A_{f_1}(\Delta) = \int_{-\infty}^{\infty} P(\phi, t) \cos(\phi) d\phi = \exp\left[-\langle \phi^{\beta}(t) \rangle / 2\right]. \tag{19}$$

When the FGPW effect is taken into account, the accumulating phase shift distribution including the FGPW effect will approximately obey Eq. (18), and the relationship between $\langle |\phi^{\beta}(t)| \rangle$ and $\langle \phi^{2}(t) \rangle$ will be approximately described by Eq. (17). The direct calculation of $\langle |\phi^{\beta}(t)| \rangle / 2$ is difficult. However, it can be calculated indirectly via Eq. (17) [33]. Based on Eq. (12), we have

$$\frac{\langle \phi^{2}(t) \rangle}{\frac{1}{\sqrt{\pi}} \Gamma\left(\frac{2}{\beta} + \frac{1}{2}\right)} = \frac{1}{\frac{1}{\sqrt{\pi}} \Gamma\left(\frac{2}{\beta} + \frac{1}{2}\right)} \left\langle \left[ \int_{0}^{t} \gamma g(t') z(t') dt' \right]^{2} \right\rangle. \tag{20}$$

For the pulsed gradient spin echo (PGSE) and the pulsed gradient stimulated-echo (PGSTE) experiments as shown in Figure 1, with constant gradient $g$, $\langle \phi^{2}(t) \rangle$ can be calculated as [13, 35, 27]

$$\langle \phi^{2}(t) \rangle = \left\langle \left[ \int_{0}^{t} \gamma g(t') z(t') dt' \right]^{2} \right\rangle = \int_{0}^{\delta} \int_{0}^{\delta} \langle z(t') z(t'') \rangle dt' dt'' + \int_{\Delta}^{\Delta+\delta} \int_{\Delta}^{\Delta+\delta} \langle z(t') z(t'') \rangle dt' dt'' - 2 \int_{0}^{\delta} \int_{\Delta}^{\Delta+\delta} \langle z(t') z(t'') \rangle dt' dt''. \tag{21}$$

From Eq. (11),

$$\langle z^{2}(t) \rangle = \frac{\Gamma\left(\frac{2}{\beta} + \frac{1}{2}\right)}{\sqrt{\pi}} 4^{\frac{2}{\beta}} D_{f_1}^{\frac{2}{\beta}} t^{\nu} , \tag{22}$$

where $v = 2\alpha/\beta$. The similar calculation of Eq. (21) including time correlation function $z(t')z(t'')$ has been derived by Kärger et al. (see Appendix A). It must be noted that Kärger et al.'s result is for subdiffusion with $\alpha < 1$, however, their results can be extended to $0 < \alpha < 2$ in a straightforward manner. Based on Kärger et al.'s result [13], it can be gotten that

$$\frac{\langle \phi^2(t) \rangle}{\frac{1}{\sqrt{\pi}}\Gamma\left(\frac{2}{\beta}+\frac{1}{2}\right)} = \frac{2\gamma^2 g^2 4^{\frac{2}{\beta}} D_{f_1}^{\frac{2}{\beta}}}{(v+1)(v+2)}\left[\frac{1}{2}(\Delta+\delta)^{v+2} + \frac{1}{2}(\Delta-\delta)^{v+2} - \Delta^{v+2} - \delta^{v+2}\right]. \tag{23}$$

By substituting Eq. (23) into Eq. (17), we have

$$\frac{\langle |\phi^\beta(t)| \rangle}{2} = \gamma^\beta g^\beta D_{f_1} \left\{ \frac{2}{(\frac{2\alpha}{\beta}+1)(\frac{2\alpha}{\beta}+2)}\left[\frac{1}{2}(\Delta+\delta)^{\frac{2\alpha}{\beta}+2} + \frac{1}{2}(\Delta-\delta)^{\frac{2\alpha}{\beta}+2} - \Delta^{\frac{2\alpha}{\beta}+2} - \delta^{\frac{2\alpha}{\beta}+2}\right]\right\}^{\frac{\beta}{2}}. \tag{24}$$

By substituting Eq. (24) into Eq. (19), we obtain

$$A_{f_1}(t) = \exp\left[-D_{f_1} b^*_{\alpha,\beta}\right]. \tag{25}$$

where

$$b^*_{\alpha,\beta} = \gamma^\beta g^\beta \left[\frac{2}{(\frac{2\alpha}{\beta}+1)(\frac{2\alpha}{\beta}+2)}\left[\frac{1}{2}(\Delta+\delta)^{\frac{2\alpha}{\beta}+2} + \frac{1}{2}(\Delta-\delta)^{\frac{2\alpha}{\beta}+2} - \Delta^{\frac{2\alpha}{\beta}+2} - \delta^{\frac{2\alpha}{\beta}+2}\right]\right]^{\frac{\beta}{2}} \tag{26}$$

under SGP approximation, Eq. (25) reduces to

$$A_{f_1}(\Delta) = \exp\left[-D_{f_1}(\gamma g \delta)^\beta \Delta^\alpha\right]. \tag{27}$$

which is exactly the same as the effective phase shift diffusion equation and instantaneous signal attenuation method result [25, 26]. When $\alpha = 1, \beta = 2$, the result agrees with the normal diffusion result $\exp\left[-D\gamma^2 g^2 \delta^2 (\Delta - \delta/3)\right]$ [11, 37]. When $\beta = 2$, Eq. (25) reduces to

$$A_{f_1}(t) = \exp\left\{-\frac{2D_{f_1}\gamma^2 g^2}{(\alpha+1)(\alpha+2)}\left[\frac{1}{2}(\Delta+\delta)^{\alpha+2} + \frac{1}{2}(\Delta-\delta)^{\alpha+2} - \Delta^{\alpha+2} - \delta^{\alpha+2}\right]\right\}, \tag{28}$$

which agrees with Kärger's results [13, 35].

## 3. Results and discussion

**Table 1.** Comparison of the signal attenuation from this method with stretched exponential function based signal attenuation obtained by other methods.

| General Fractional diffusion $\{0 < \alpha, \beta \leq 2\}$ |
|---|
| This method: $\exp[-D_{f_1}(\gamma g \delta)^\beta \Delta^\alpha]$, $\exp[-D_{f_1} b^*_{\alpha,\beta}]$, <br> $b^*_{\alpha,\beta} = \gamma^\beta g^\beta \left[\dfrac{2}{(\nu+1)(\nu+2)}\left[\dfrac{1}{2}(\Delta+\delta)^{\nu+2} + \dfrac{1}{2}(\Delta-\delta)^{\nu+2} - \Delta^{\nu+2} - \delta^{\nu+2}\right]\right]^{\frac{\beta}{2}}$, $\nu = \dfrac{2\alpha}{\beta}$ |
| Both EPSDE method [25] and ISA method [26] got an identical result, which is $\exp\left[-D_{f_1}\int_0^\tau K^\beta(t)dt^\alpha\right]$. |
| Space-fractional diffusion $\{\alpha = 1, 0 < \beta \leq 2\}$ |
| This method: $\exp[-\gamma^\beta g^\beta D_{f_1} C_{1,\beta}(\delta, \Delta)]$, <br> $C_{1,\beta}(\delta,\Delta) = \left[\dfrac{2}{(\frac{2}{\beta}+1)(\frac{2}{\beta}+2)}\left[\dfrac{1}{2}(\Delta+\delta)^{\frac{2}{\beta}+2} + \dfrac{1}{2}(\Delta-\delta)^{\frac{2}{\beta}+2} - \Delta^{\frac{2}{\beta}+2} - \delta^{\frac{2}{\beta}+2}\right]\right]^{\frac{\beta}{2}}$ |
| Other methods: <br> $\exp\left[-D_{f_1}\gamma^\beta g^\beta \delta^\beta(\Delta - \dfrac{\beta-1}{\beta+1}\delta)\right]$ from EPSDE [19] and ISA [20] methods <br> $\exp\left[-D\mu^{2(\beta''/2-1)}(\gamma g \delta)^{\beta''}(\Delta - \dfrac{\beta''-1}{\beta''+1}\delta)\right]$ from Magin et al. [21], $\exp(-K^\mu q^\mu t)$ [18] |
| Time-fractional diffusion $\{0 < \alpha \leq 2, \beta = 2\}$ |
| $C'_{\alpha,2}(\delta,\Delta) = \gamma^2 g^2 \left\{\dfrac{\alpha\delta^{\alpha+2}}{\alpha+2} + \delta^2(\Delta^\alpha - \delta^\alpha) + \dfrac{2}{(\alpha+1)(\alpha+2)}\left[(\Delta+\delta)^{2+\alpha} - \Delta^{2+\alpha}\right] - \dfrac{2}{(\alpha+1)}\Delta^{1+\alpha}\delta - \Delta^\alpha\delta^2\right\}$ <br> $\exp\{-D_{f_1}\gamma^2 g^2 C'_{\alpha,2}(\delta,\Delta)\} = \begin{cases}\exp(-D_{f_1}\gamma^2 g^2 \delta^2(\Delta - \delta/3), \alpha = 1 \\ \exp\left[-D_{f_1}\gamma^2 g^2 \delta^{\alpha+2}\dfrac{4(2^\alpha - 1)}{(\alpha+1)(\alpha+2)}\right], \Delta = \delta \\ \exp(-D_{f_1}\gamma^2 g^2 \delta^2 \Delta^\alpha), \text{if } \delta \text{ is neglected.}\end{cases}$ |
| Other methods: <br> 1. Kärger et al.'s: $\exp\{-D\gamma^2 g^2 C'_{\alpha,2}(\delta,\Delta)\}$. [13,35] <br> 2. EPSDE [25] and ISA methods [26]: $\exp(-T) = \begin{cases}\exp(-D_{f_1}\gamma^2 g^2 \delta^2(\Delta - \delta/3), \alpha = 1 \\ \exp\left[-D_{f_1}\gamma^2 g^2 \delta^{\alpha+2}\dfrac{4(2^{\alpha+1} - \alpha - 2)}{(\alpha+1)(\alpha+2)}\right], \Delta = \delta \\ \exp(-D_{f_1}\gamma^2 g^2 \delta^2 \Delta^\alpha), \text{if } \delta \text{ is neglected.}\end{cases}$ <br> $T = D_{f_1}\gamma^2 g^2 \left\{\dfrac{\alpha\delta^{\alpha+2}}{\alpha+2} + \delta^2(\Delta^\alpha - \delta^\alpha) + \dfrac{2}{(\alpha+1)(\alpha+2)}\left[(\Delta+\delta)^{2+\alpha} - \Delta^{2+\alpha}\right] - \dfrac{2}{(\alpha+1)}\Delta^{1+\alpha}\delta - \Delta^\alpha\delta^2\right\}$ <br> 3. $\exp\left[-D\gamma^2 g^2 \delta^2(\Delta - \delta/3)\right]$, free normal diffusion [11, 37] <br> 4. $\exp(-bD)^\alpha$ experimental, Bennett et al. [16] <br> 5. $\exp\{-D_\nu g^2 t_e^{3-\nu}/[2^{2-\nu}(3-\nu)]\}, \langle z^2 \rangle \propto t^{1-\nu}$, from Wisdom et al. [39] <br> 6. $\exp(-K_\alpha q^2 t^\alpha)$ [18] |

A modified Gaussian phase distribution method has been developed for calculating the accumulating phase shift distribution and the signal attenuation in PFG experiments for fractional diffusion based on fractal derivative model. The signal attenuation expression can describe all three types of fractional diffusion and can handle the finite gradient pulse width (FGPW) effect that is important in MRI. While many other results can only deal with one or two types of fractional diffusions and do not treat FGPW effect. Table 1 shows the comparison of the results of this method with other results in the literature that give a stretched exponential function based signal attenuation.

From Table 1, the approximation expression obtained in this paper is consistent with the reported results in the literature. In the time fractional diffusion, the result obtained here agrees with Kärger et al.'s result. When $\Delta = \delta$, namely $\delta/(\Delta-\delta)$ is infinitely large due to $\Delta-\delta = 0$, the time fractional diffusion signal attenuation expression Eq. (28) reduces to $\exp\left[-D_{f_1}\gamma^2 g^2 \delta^{\alpha+2} \frac{4(2^\alpha - 1)}{(\alpha+1)(\alpha+2)}\right]$, which is close to $\exp\left[-D_{f_1}\gamma^2 g^2 \delta^{\alpha+2} \frac{4(2^{\alpha+1} - \alpha - 2)}{(\alpha+1)(\alpha+2)}\right]$ from the effective phase shift diffusion equation method [25] and instantaneous signal attenuation method [26] within a certain range of $\alpha$. Both of these signal attenuation expressions depend on $\delta^{\alpha+2}$. The differences between the constants $\frac{4(2^\alpha - 1)}{(\alpha+1)(\alpha+2)}$ and $\frac{4(2^{\alpha+1} - \alpha - 2)}{(\alpha+1)(\alpha+2)}$ at different $\alpha$ values are calculated and shown in Figure 2. From Figure 2, the maximum difference is within 20 % between the two expressions for $0.5 < \alpha < 1.5$. When $\alpha$ approaches 1, the differences become smaller and smaller. This difference will make the diffusion coefficient obtained by this method different from that of the effective phase shift diffusion equation and instantaneous signal attenuation methods when $\Delta = \delta$. Such a difference becomes smaller as $\delta : \Delta$ become smaller, since under SGP approximation, all the attenuation expressions from this method, the effective phase shift diffusion equation method, and the instantaneous signal attenuation method reduce to $\exp(-D_{f_1}\gamma^2 g^2 \delta^2 \Delta^\alpha)$. Therefore, in many practical PFG NMR experiments such as those performed in high field NMR spectrometers, as $\delta/(\Delta-\delta)$ is not too large, the total difference among this method, the effective phase shift diffusion equation method and the instantaneous signal attenuation method is small.

Figure 3 shows the comparison of this method, the effective phase shift diffusion equation method and instantaneous signal attenuation method at different $\alpha$ and $\beta$ combinations with various $\delta/(\Delta-\delta)$ ratios. When the $\delta/(\Delta-\delta)$ ratio decreases, the difference between the result of this method and that of the effective phase shift diffusion equation and instantaneous signal attenuation methods decreases. For a large $\delta/(\Delta-\delta)$ ratio such as $\delta/(\Delta-\delta) = 1.5$, when $0.75 < \alpha < 1.5$ and $0.1 < \beta \leq 2$, most of the values of $\ln[A_{Eq.(20)}(t)]/\ln[A_{EPSDE}(t)]$ are within the range from 0.75 to 1.2. For a small $\delta/(\Delta-\delta)$ ratio such a $\delta/(\Delta-\delta) = 0.2$, when $0.75 < \alpha < 1.5$ and $0.1 < \beta \leq 2$, most of the values of $\ln[A_{Eq.(20)}(t)]/\ln[A_{EPSDE}(t)]$ are within the range of 0.85 to 1.15. For the narrow gradient pulse regime, such as $\delta/(\Delta-\delta) = 10^{-6}$, the difference among different methods is negligible as all the methods reduce to $\exp\left[-D_{f_1}(\gamma g \delta)^\beta \Delta^\alpha\right]$. When the $\beta$ value is fixed, the value of $\ln[A_{Eq.(20)}(t)]/\ln[A_{EPSDE}(t)]$ increases when the $\alpha$ value decreases, while, when the $\alpha$ value is fixed, the value of $\ln[A_{Eq.(20)}(t)]/\ln[A_{EPSDE}(t)]$ does not monotonically increase or decrease inside the range of $0.1 < \beta \leq 2$. This method is convenient as it can be used to estimate the attenuation easily, which is very helpful for setting PFG experiment. The possible origin of the difference is that the phase distribution may deviate from Eq. (18) and the mean $\beta$-th power of the phase shift does not obey Eq. (17). Additionally, the neglecting of $K(t)$ in the effective phase shift diffusion equation or instantaneous signal attenuation method may underestimate the signal attenuation at small $\alpha$ and large $t$. At small $\alpha$ and large $t$, the increasement of the mean $\beta$-th power displacement $D_{f_1}\alpha t^{\alpha-1}dt$ during $dt$ becomes smaller, the change of $K(t)$ during $dt$ may become comparable to $D_{f_1}\alpha t^{\alpha-1}dt$ and may not be negligible. Currently, the exact origin of the difference is still not clear, reminding us that more concentrated efforts are needed in this direction.

The derivative parameters $\alpha$ and $\beta$ have been used as contrast parameters for MRI [23]. Additionally, Palombo et al. have used NMR to measure parameters $\alpha$ and $\beta$ by phenomenological signal attenuation formalisms for time-fractional diffusion and space-fractional diffusion that are based on a continuous time random walk [18]. Based on the signal attenuation expression obtained in this paper, Eq. (25), a new set of formalisms directly determining parameters $\alpha$ and $\beta$ can be derived. From Eqs. (25) and (26), when the time constant $\delta$ and $\Delta$ are fixed, we have

$$\ln[\ln A(0) - \ln A(t)] = \ln[\gamma^\beta D_{f_1} C_{\alpha,\beta}(\delta,\Delta)] + \beta \ln(g), \tag{29}$$

where $C_{\alpha,\beta}(\delta,\Delta)$ is

$$C_{\alpha,\beta}(\delta,\Delta) = \left[\frac{2}{(\frac{2\alpha}{\beta}+1)(\frac{2\alpha}{\beta}+2)}\left[\frac{1}{2}(\Delta+\delta)^{\frac{2\alpha}{\beta}+2} + \frac{1}{2}(\Delta-\delta)^{\frac{2\alpha}{\beta}+2} - \Delta^{\frac{2\alpha}{\beta}+2} - \delta^{\frac{2\alpha}{\beta}+2}\right]\right]^{\frac{\beta}{2}}. \tag{30}$$

From Eqs. (25 and (26), we can also have

$$\ln[\ln A(0) - \ln A(t)] = \begin{cases} \ln[\gamma^\beta g^\beta D_{f_1} C_2] + (\alpha+\beta)\ln(\delta), & \text{when } \Delta = \delta \\ \ln[\gamma^\beta g^\beta D_{f_1}] + \alpha \ln(\Delta), & \text{when } \delta \ll \Delta \end{cases}, \tag{31}$$

where $C_2$ is

$$C_2 = \left[\frac{4\delta^{\frac{2\alpha}{\beta}+2}}{(\frac{2\alpha}{\beta}+1)(\frac{2\alpha}{\beta}+2)}\left[2^{\frac{2\alpha}{\beta}+1}-1\right]\right]^{\frac{\beta}{2}}. \tag{32}$$

Equations similar to Eqs. (29) and (31) have been given in literature [26, 33]. Eqs. (29) and (31) may be used to determine the fractional time and space derivative parameters $\alpha$ and $\beta$ in PFG fractional diffusion experiments.

This method is distinct from other methods such as the effective phase shift diffusion equation method [25] and instantaneous method [26]. In this method, it is assumed that the phase distribution including the FGPW effect can be approximately treated as the same type of phase distribution that is obtained from the fractional diffusion equation under SGP approximation, and the approximate phase distribution is determined by the mean $\beta$-th power of the phase displacement that can be calculated by Eqs. (17) and (23). Therefore, this is a type of method similar the GPD method for normal diffusion. While, the effective phase shift diffusion equation method obtains the phase distribution including FGPW effect by solving diffusion equation directly; the instantaneous signal attenuation method approximately treats the total signal attenuation as an accumulating effect of the signal attenuation of every small interval during PFG diffusion process.

There are similarities and differences between results of this paper and those obtained in reference [33]. Both results use a similar idea, modified from the traditional Gaussian phase distribution approximation method. However, there are significant differences between the fractal derivative model used here and the fractional derivative model used in reference [33]. These two derivative models have different diffusion equations and probability distribution functions. Additionally, the signal attenuation obtained from the fractal derivative model in this paper is a stretched exponential function based attenuation, while, the signal attenuation based on the fractional derivative model is a Mittag-Leffler function based signal attenuation, which is $A(t) = E_\alpha\left(-D_{f_2} b^*_{\alpha,\beta}\right)$ where $D_{f_2}$ is a fractional diffusion coefficient based on the fractional derivative, and $b^*_{\alpha,\beta}$ is the same as that defined according to Eq. (26). This is the same as that in $A(t) = \exp\left(-D_{f_1} b^*_{\alpha,\beta}\right)$ obtained in this paper. The Mittag-Leffler function based attenuation and the stretched exponential function based attenuation are equivalent at small attenuation, but deviate from each other at larger attenuations. Moreover, from the fractal derivative model, the mean $\eta$-th power of the displacement $\langle|\phi^\eta(\Delta)|\rangle$ can be calculated according to Eq. (16), but it is still difficult to calculate in the fractional derivative model. Currently, there are still many open problems in the PFG fractional diffusion, such as the Mittag-Leffler function based attenuation for supper diffusion giving a negative signal attenuation at a certain range of the value of $D_{f_2}\int_0^\tau K^\beta(t)dt^\alpha$ that is hard to explain. The stretched exponential function will not give a negative PFG signal attenuation. Therefore, there are similarities and differences in these two derivative models. It was pointed out in reference [30] that both models can depict a variety of diffusion processes but it is still unclear of their particular application areas. The fractal derivatives are local operators, while the fractional derivatives are global operators [30].

The stretched exponential function can be represented by a series expansion of monoexponential function (see Appendix B). However, the expansion in Appendix B shows that the distributed diffusion center $\Lambda$ (see Eq. (B.6) in Appendix B) is a time-dependent variable. This is a noticeable difference between the fractional diffusion result and the commonly used multiple component expansion methods such as bi-exponential model [16] with a fixed distributed diffusion center.

The modified Gaussian phase distribution approximation based on fractal derivative provides a set of useful formalisms to analyze signal attenuation and to determine parameter $\alpha$ and $\beta$ for PFG fractional diffusion experiments. The method here gives an alternative way to analyze PFG fractional diffusion. Researchers may benefit from the results here as they can choose one suitable type of fractional diffusions among these three kinds of fractional diffusions for their experimental systems. Additionally, the results here can treat the FGPW effect, which can improve the accuracy of interpretation of these PFG experiments where the gradient pulse is not infinitely narrow, particularly in PFG MRI experiments. This method could be improved and extended to various PFG fractional diffusion problems such as restricted fractional diffusion including the FGPW effect in future work.

**Appendix A. Kärger et al.'s correlation expression [13, 35].**

For an anomalous diffusion with $\langle z^2(t) \rangle = 2Dt^\alpha$, Kärger et al. shown that the mean correlated displacement $\langle z(t')z(t' \pm \theta) \rangle$ can be described by [13, 35]

$$\langle z(t')z(t' \pm \theta) \rangle = D\left[(t' \pm \theta)^\alpha + t'^\alpha - \theta^\alpha \right]. \tag{A.1}$$

where $z(t')$ and $z(t' \pm \theta)$ are the positions at time $t'$ and $t' \pm \theta$, respectively. The reader can be further referred to reference [13,35] for more detail information.

**Appendix B. Expansion of stretched exponential attenuation.**

The stretched exponential function has broad applications. For instance, the Kohlrausch-Williams-Watts (KWW) function is often used to describe the correlation time distribution in polymer systems. The commonly used KWW function can be written as $\exp[-(\frac{t}{\tau_p})^\sigma]$, where $\tau_p$ is the center correlation time, and $\sigma$ is the breadth parameter. In PFG diffusion study, the signal attenuation expressions similar to KWW function have been used to describe signal attenuation [40]. Bennett et al.'s reported a phenomenological PFG signal attenuation expression $\exp\left[-(bDDC)^\alpha\right]$ (DDC is the distributed diffusion coefficient) to describe water diffusion in brain and pointed out that it can be expanded into $\int_0^\infty \rho(D)\exp(-bD)dD$. Here, it will be illustrated that the PFG signal attenuation expression Eq. (25) can be expanded similarly and the expansion coefficient can be obtained. It has been shown that the KWW function can be expanded into a continuous distribution as [41, 42]

$$\exp[-(\frac{t}{\tau_p})^\sigma] = \int_0^\infty \rho(\tau)\exp(-\frac{t}{\tau})d\tau = \int_{-\infty}^\infty d(\log\tau)G(\log\tau)\exp(-\frac{t}{\tau}), \tag{B.1}$$

or a discrete sum as [43,44,45]

$$\exp[-(\frac{t}{\tau_p})^\sigma] = \sum_j P_j \exp(-\frac{t}{\tau_j}) \tag{B.2}$$

with $P_j = G(\log\tau_j)\log\left(\frac{\tau_{j+1}}{\tau_j}\right)$, where $G(\log\tau)$ is defined as [43]

$$G(\log \tau) = [\tau \rho_\sigma(\tau)/\log e], \qquad (B.3)$$

where the associated distribution function $\rho_\sigma(\tau)$ is [43]

$$\rho_\sigma(\tau) = \frac{1}{\pi \tau} \sum_{k=0}^{\infty} (-1)^{k+1} \frac{\Gamma(1+k\sigma)}{k!} \left(\frac{\tau}{\tau_p}\right)^{k\sigma} \sin(\pi k \sigma). \qquad (B.4)$$

In the numerical evaluation, the convergence of Eq. (B.3) can be slow [43]. In polymer relaxation studies, there are other ways to expand $\rho_\sigma(\tau)$ such as the lognormal expansion [44,45,46] that may give fast convergence in computation. For interested readers, refer to references [44,45,46] for those detail expansions.

PFG signal attenuation expression Eq. (25) can be rewritten as

$$A_{f_1}(t) = \exp\left[-D_{f_1} b^*_{\alpha,\beta}\right] = \exp\left\{-\left[\frac{\left(D_{f_1} b^*_{\alpha,\beta}\right)^{\frac{2}{\beta}}}{b}\right]^{\frac{\beta}{2}} b\right\} = \exp\left[-\left[\frac{b}{\Lambda}\right]^{\frac{\beta}{2}}\right], \qquad (B.5)$$

where $b = \gamma^2 g^2 \delta^2 \left(\Delta - \frac{\delta}{3}\right)$, and $\Lambda$ is

$$\Lambda = \frac{b}{\left(D_{f_1} b^*_{\alpha,\beta}\right)^{\frac{2}{\beta}}} = \frac{\gamma^2 g^2 \delta^2 \left(\Delta - \frac{\delta}{3}\right)}{\left[D_{f_1} \gamma^\beta g^\beta C_{\alpha,\beta}(\delta,\Delta)\right]^{\frac{2}{\beta}}}$$

$$= \frac{\delta^2 \left(\Delta - \frac{\delta}{3}\right)}{\dfrac{2 D_{f_1}^{\frac{2}{\beta}}}{(\frac{2\alpha}{\beta}+1)(\frac{2\alpha}{\beta}+2)} \left[\frac{1}{2}(\Delta+\delta)^{\frac{2\alpha}{\beta}+2} + \frac{1}{2}(\Delta-\delta)^{\frac{2\alpha}{\beta}+2} - \Delta^{\frac{2\alpha}{\beta}+2} - \delta^{\frac{2\alpha}{\beta}+2}\right]}, \qquad (B.6)$$

which is a time dependent variable. Under SGP approximation, Eq. (B.6) reduces to

$$\Lambda = \frac{(\frac{2\alpha}{\beta}+1)(\frac{2\alpha}{\beta}+2)}{2 D_{f_1}^{\frac{2}{\beta}}} \Delta^{1-\frac{2\alpha}{\beta}}, \qquad (B.7)$$

which is clearly a time-dependent variable except when $2\alpha/\beta$ equals 1. The PFG signal attenuation expression, Eq. (B.5) is similar to the KWW function, therefore, it can be expanded similarly according to Eq. (B.2) as

$$\exp\left[-\left(\frac{b}{\Lambda}\right)^{\frac{\beta}{2}}\right] = \sum_j P_j \exp(-D_j b), \tag{B.8}$$

where

$$P_j = \left[D_j \rho_{\beta/2}(D_j)/\log e\right]\log\left(\frac{D_{j+1}}{D_j}\right), \tag{B.9}$$

where $\rho_{\beta/2}(D)$ can be obtained from Eq. (B.3)

$$\rho_{\beta/2}(D) = \frac{1}{\pi D}\sum_{k=0}^{\infty}(-1)^{k+1}\frac{\Gamma(1+k\beta/2)}{k!}\left(\frac{D}{\Lambda}\right)^{k\beta/2}\sin(\pi k\beta/2). \tag{B.10}$$

The time dependent $\Lambda$ determines that the center of the distribution of diffusion coefficient is time-dependent. From Eq. (B.7), the PFG diffusion with a stretched exponential signal attenuation described by Eq. (25) can be viewed as equivalent to the sum of a distribution of monoexponential attenuation with a time dependent distributed diffusion coefficient center. A value of parameter $\beta/2$ in Eq. (B.8) near 1 means narrow distribution, while a lower value of $\beta/2$ means broader space distribution.

Besides $\exp\left[-(b/\Lambda)^{\beta/2}\right]$, the PFG signal attenuation expression may be rewritten into other forms belonging to KWW function. For example, the time fractional signal attenuation expression $\exp(-D_{f_1}\gamma^2 g^2\delta^2\Delta^\alpha)$ can be rewritten as $\exp\left\{-\left[(D_{f_1}\gamma^2 g^2\delta^2)^{1/\alpha} b\right]^\alpha\right\}$, but it may not be useful since practical PFG experiments that are often carried out by varying gradient intensities cannot determine $(D_{f_1}\gamma^2 g^2\delta^2)^{1/\alpha}$, a gradient intensity dependent term.

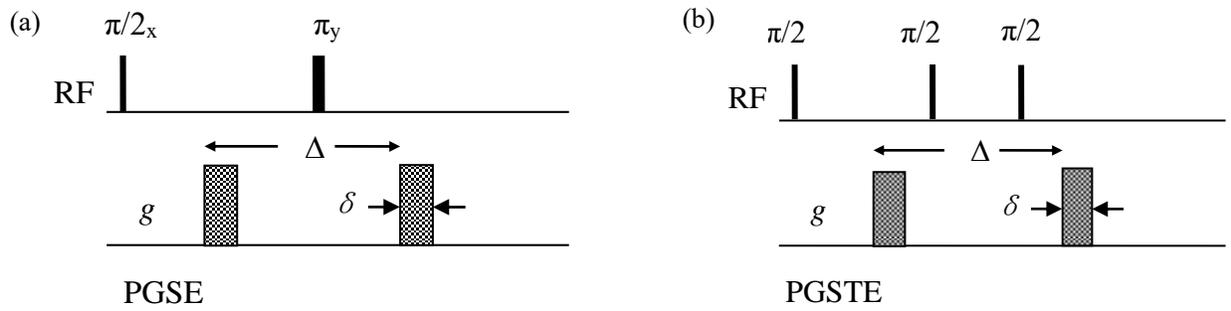

**FIG. 1** Two typical PFG diffusion sequence: (a) PGSE pulse sequences, (b) PGSTE pulse sequence.

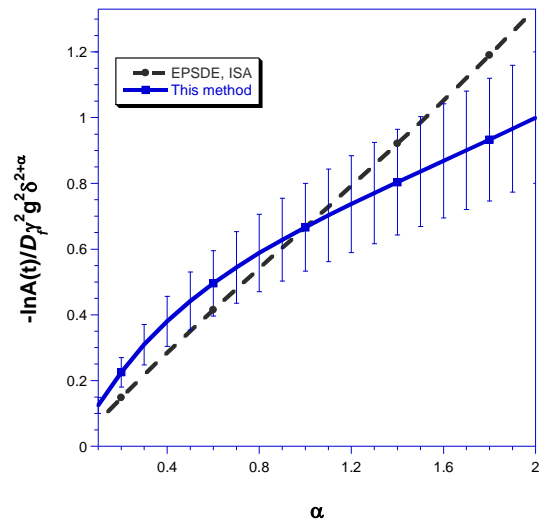

**FIG. 2** Comparison of $-\ln A(t)/D_f \gamma^2 g^2 \delta^{\alpha+2}$ in the time fractional diffusion signal attenuation expression of Eq. (28) and that obtained from other method at different $\alpha$ values.

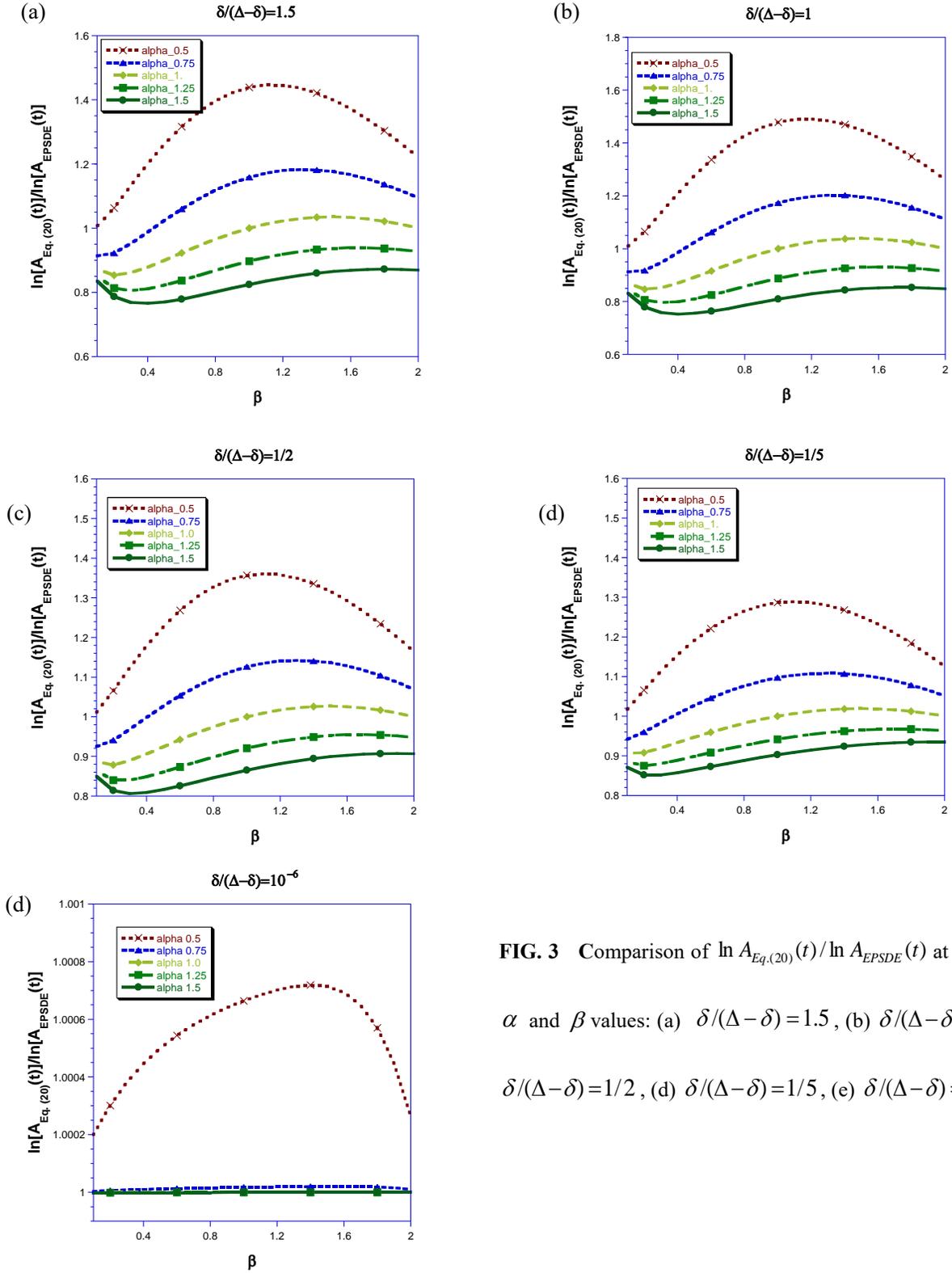

**FIG. 3** Comparison of $\ln A_{Eq.(20)}(t)/\ln A_{EPSDE}(t)$ at different $\alpha$ and $\beta$ values: (a) $\delta/(\Delta-\delta)=1.5$, (b) $\delta/(\Delta-\delta)=1$, (c) $\delta/(\Delta-\delta)=1/2$, (d) $\delta/(\Delta-\delta)=1/5$, (e) $\delta/(\Delta-\delta)=10^{-6}$.

**References:**

1. R. Metzler, J. Klafter, The random walk's guide to anomalous diffusion: a fractional dynamics approach, Phys. Rep. 339 (2000) 1–77.
2. R. Metzler, E. Barkai, J. Klafter Anomalous transport in disordered systems under the influence of external fields, Physica A, 266 (1999) 343–350.
3. E.K. Lenzi, M.A.F. dos Santos, D.S. Vieira, R.S. Zola, H.V. Ribeiro, Solutions for a sorption process governed by a fractional diffusion equation, Physica A 447 (2016) 467–481.
4. I. M. Sokolov, Models of anomalous diffusion in crowded environments, Soft Matter, 8 (2012) 9043-9052.
5. S. Burov, J.-H. Jeon, R. Metzler, E. Barkai, Single particle tracking in systems showing anomalous diffusion: the role of weak ergodicity breaking, Phys. Chem. Chem. Phys., 13 (2011) 1800–1812.
6. R. Metzler, J.-H. Jeon, A. G. Cherstvy, E. Barka, Anomalous diffusion models and their properties: non-stationarity, non-ergodicity, and ageing at the centenary of single particle tracking, Phys.Chem.Chem.Phys., 16 (2014) 24128
7. Y. Meroza, I. M. Sokolov, A toolbox for determining subdiffusive mechanisms, Phys. Reports, 573 (2015) 1–29.
8. E.L. Hahn, Spin Echoes, Phys. Rev. 80 (1950) 580.
9. D. W. McCall, D. C. Douglass, E. W. Anderson, B. Bunsenges, Physik. Chem. 67 (1963) 336.
10. E. O. Stejskal. J. E. Tanner. J. Chem. Phys. 42 (1965) 288; doi: 10.1063/1.1695690.
11. P. Callaghan, Translational Dynamics and Magnetic Resonance: Principles of Pulsed Gradient Spin Echo NMR, Oxford University Press, 2011.
12. T. Zavada, N. Südland, R. Kimmich, T.F. Nonnenmacher, Propagator representation of anomalous diffusion: The orientational structure factor formalism in NMR, Phys. Rev. E 60 (1999) 1292-1298.
13. J. Kärger, H. Pfeifer, G. Vojta, Time correlation during anomalous diffusion in fractal systems and signal attenuation in NMR field-gradient spectroscopy, Phys. Rev. A 37 (11) (1988) 4514–4517.
14. N. Fatkullin, R. Kimmich, Theory of field-gradient NMR diffusometry of polymer segment displacements in the tube-reptation model, PHYS. REV. E 52 (1995) 3273-3276
15. R. A. Damion, K.J. Packer, Predictions for Pulsed-Field-Gradient NMR Experiments of Diffusion in Fractal Spaces, Proceedings: Mathematical, Physical and Engineering Sciences, 453 (1997) 205-211.
16. K.M. Bennett, K.M. Schmainda, R.T. Bennett, D.B. Rowe, H. Lu, J.S. Hyde, Characterization of continuously distributed cortical water diffusion rates with a stretched-exponential model, Magn. Reson. Med. 50 (2003) 727–734.
17. K.M. Bennett, J.S. Hyde, K.M. Schmainda, Water diffusion heterogeneity index in the human brain is insensitive to the orientation of applied magnetic field gradients, Magn. Reson. Med. 56 (2006) 235–239.
18. M. Palombo, A. Gabrielli, S. D. Santis, C. Cametti, G. Ruocco, S. Capuani, Spatio-temporal anomalous diffusion in heterogeneous media by nuclear magnetic resonance, J. Chem. Phys. 135 (2011) 034504.
19. S. Capuani, M. Palombo, A. Gabrielli, A. Orlandi, B. Maraviglia, F. S. Pastore, Spatio-temporal anomalous diffusion imaging: results in controlled phantoms and in excised human meningiomas, Magn. Reson. Imaging, 31 (2013) 359.
20. E. Özarslan, P.J. Basser, T.M. Shepherd, P.E. Thelwall, B.C. Vemuri, S.J. Blackband, Observation of anomalous diffusion in excised tissue by characterizing the diffusion-time dependence of the MR signal, J. Magn. Reson. 183 (2006) 315–323.

21. R.L. Magin, O. Abdullah, D. Baleanu, X.J. Zhou, Anomalous diffusion expressed through fractional order differential operators in the Bloch–Torrey equation, J. Magn. Reson. 190 (2008) 255–270.

22. A. Hanyga, M. Seredyńska, Anisotropy in high-resolution diffusion-weighted MRI and anomalous diffusion, Journal of Magnetic Resonance 220 (2012) 85–93.

23. C. Ingo, R.L. Magin, L. Colon-Perez, W. Triplett, T.H. Mareci, On random walks and entropy in diffusion-weighted magnetic resonance imaging studies of neural tissue, Magn. Reson. Med. 71 (2014) 617–627.

24. Grinberg F, Ciobanu L, Farrher E, Shah NJ (2012) Diffusion kurtosis imaging and log-normal distribution function imaging enhance the visualization of lesions in animal stroke models. NMR Biomed. 25:1295–1304.

25. G. Lin, An effective phase shift diffusion equation method for analysis of PFG normal and fractional diffusions, J. Magn. Reson. 259 (2015) 232-240.

26. G. Lin, Instantaneous Signal Attenuation Method for Analysis of PFG Fractional Diffusions, J. Magn. Reson.269 (2016) 36-49.

27. D. Le Bihan, M. Iima, Diffusion Magnetic Resonance Imaging: What Water Tells Us about Biological Tissues. PLoS Biol. 13 (7) (2015) e1002203. doi:10.1371/journal.pbio.1002203

28. F. Grinberg, E. Farrher, L. Ciobanu, F. Geffroy, D. Le Bihan, N. J. Shah. Non-Gaussian Diffusion Imaging for Enhanced Contrast of Brain Tissue Affected by Ischemic Stroke. PLoS ONE. 9 (2) (2014) e89225. doi: 10.1371/journal.pone.0089225 PMID: 24586610.

29. W. Chen, Time space fabric underlying anomalous diffusion, Chaos, Solitons Fract. 28 (2006) 923.

30. W. Chen, H. Sun, X. Zhang, D. Korošak, Anomalous diffusion modeling by fractal and fractional derivatives, Comput. Math. Appl. 59 (5) (2010) 1754–1758.

31. F. Mainardi, Y. Luchko and G. Pagnini, The fundamental solution of the space-time-fractional diffusion equation, Fract. Calc. Appl. Anal. 4 (2001) 153.

32. R. Gorenflo, F. Mainardi, Fractional Diffusion Processes: Probability Distributions and Continuous Time Random Walk, Springer Lecture Notes in Physics, No 621, Berlin 2003, pp. 148-166

33. G. Lin, Analyzing Signal Attenuation in PFG Anomalous Diffusion via a Non-Gaussian Phase Distribution Approximation Approach by fractional derivatives, Submitted to The Journal of Chemical Physics.

34. H. Cao, G. Lin, A. Jones, Anomalous penetrant diffusion as a probe of the local structure in a blend of poly(ethylene oxide) and poly(methyl methacrylate), J. Polym. Sci. Part B: Polym. Phys. 42 (2004) 1053–1067.

35. R. Kimmich, NMR: Tomography, Diffusometry, Relaxometry, Springer-Verlag, Heidelberg, 1997.

36. H.G. Sun, M. M. Meerschaert, Y. Zhang, J. Zhu, W. Chen. A fractal Richards' equation to capture the non-Boltzmann scaling of water transport in unsaturated media. Advances in Water Resources, 52 (2013) 292-295.

37. W. S. Price, Pulsed-field gradient nuclear magnetic resonance as a tool for studying translational diffusion: Part 1. Basic theory, Concepts Magn. Reson. 9 (1997) 299.

38. G. Lin, Analyzing the special PFG signal attenuation behavior of intermolecular MQC via the effective phase shift diffusion equation method, J. Chem. Phys. 143 (2015) 164202.

39. A. Widom, H.J. Chen, Fractal Brownian motion and nuclear spin echoes, J. Phys. A: Math. Gen. 28 (1995) 1243–1247.

40. M. Kopf, C. Corinth, O. Haferkamp, T.F. Nonnenmacher, Anomolous diffusion of water in biological tissues, Biophys J 70 (1996) 2950–2958.

41. C. P. Lindsey, G. D. Patterson, Detailed comparison of the Williams-Watts and Cole-Davidson functions, J. Chem. Phys. 73 (1980) 3348–3357

42. J.I. Kaplan, A.N. Garroway, Homogeneous and inhomogeneous distributions of correlation times. Lineshapes for chemical exchange, J. Magn. Reson., 49 (1982) 464-475.

43．E.W. Montroll, J.T. Bendler, On Lévy (or stable) distributions and the Williams-Watts model of dielectric relaxation, J. Stat. Phys. 34 (1) (1984) 129-162.

44．J. T. Bendler and M. F. Shlesinger, in Studies in Statistical Mechanics, Vol. 12, M. F. Shlesinger and G. H. Weiss, Eds., North-Holland, New York (1985).

45. A.K. Roy, A.A. Jones, P.T. Inglefield, 1986Phenylene ring dynamics in solid polycarbonate: an extensive probe by carbon-13 solid-state NMR line-shape studies at two field strengths, Macromolecules, 19 (5) (1986) 1356-1362.